\documentclass[sigconf]{acmart}

\AtBeginDocument{%
  \providecommand\BibTeX{{%
    \normalfont B\kern-0.5em{\scshape i\kern-0.25em b}\kern-0.8em\TeX}}}

\setcopyright{acmlicensed}
\copyrightyear{2018}
\acmYear{2018}
\acmDOI{XXXXXXX.XXXXXXX}

\acmConference[MSR 2026]{MSR '26: Proceedings of the 23rd International Conference on Mining Software Repositories}{April 2026}{Rio de Janeiro, Brazil}

\acmISBN{978-1-4503-XXXX-X/18/06}

\usepackage{graphicx} 
\usepackage{url}
\usepackage{booktabs}
\usepackage{multirow}
\usepackage{subcaption}
\usepackage{booktabs, multirow, tabularx, ragged2e, makecell}
\usepackage{pdfpages}
\usepackage{array} 
\usepackage[most]{tcolorbox}
\usepackage{soul}
\usepackage{framed}

\newenvironment{customframe}
  {%
   \MakeFramed{\advance\hsize-\width \FrameRestore}}%
  {\endMakeFramed}

\begin{document}

\title{On Autopilot? An Empirical Study of Human–AI Teaming and Review Practices in Open Source}
\author{Haoyu Gao}
\affiliation{
 \institution{The University of Melbourne}
  \country{Victoria, Australia}
 }
\email{haoyug1@student.unimelb.edu.au}
\author{Peerachai Banyongrakkul}
\affiliation{
\institution{The University of Melbourne}
\country{Victoria, Australia}
}
\email{pbanyongrakkul@student.unimelb.edu.au}
\author{Hao Guan}
\affiliation{
\institution{The University of Melbourne}
\country{Victoria, Australia}
}
\email{guan.h@student.unimelb.edu.au}

\author{Mansooreh Zahedi}
\affiliation{
\institution{The University of Melbourne}
\country{Victoria, Australia}
}
\email{mansooreh.zahedi@unimelb.edu.au}
\author{Christoph Treude}
\affiliation{
\institution{Singapore Management University}
\country{Singapore}
}
\email{ctreude@smu.edu.sg}

\begin{abstract}
Large Language Models (LLMs) increasingly automate software engineering tasks. While recent studies highlight the accelerated adoption of ``AI as a teammate'' in Open Source Software (OSS), developer interaction patterns remain under-explored. In this work, we investigated project-level guidelines and developers' interactions with AI-assisted pull requests (PRs) by expanding the AIDev dataset to include finer-grained contributor code ownership and a comparative baseline of human-created PRs. We found that over 67.5\% of AI-co-authored PRs originate from contributors without prior code ownership. Despite this, the majority of repositories lack guidelines for AI-coding agent usage. Notably, we observed a distinct interaction pattern: AI-co-authored PRs are merged significantly faster with minimal feedback. In contrast to human-created PRs where non-owner developers receive the most feedback, AI-co-authored PRs from non-owners receive the least, with approximately 80\% merged without any explicit review. Finally, we discuss implications for developers and researchers.

\end{abstract}




\received{20 February 2007}
\received[revised]{12 March 2009}
\received[accepted]{5 June 2009}

\maketitle

\section{Introduction}
\label{sec:introduction}

LLM-powered coding agents are widely adopted across various software development tasks, including code review~\cite{lin2025codereviewqa}, code generation~\cite{fakhoury2024llm}, and program repair~\cite{le2024semantic}. This paradigm shift enables these coding agents to take on significant roles as ``coding teammates'' within open source projects. Recently, Li et al.~\cite{li2025rise} curated a large-scale OSS pull request dataset capturing two distinct behaviours: developers delegating tasks entirely to AI agents (AI-only PRs) and developers co-authoring contributions alongside AI agents (Human+AI PRs).

Despite the increasing adoption of AI coding agents in OSS development, the ways in which developers prepare for and interact with these agents remain relatively understudied. Given the rapid expansion of AI-assisted coding, effective governance and monitoring have become increasingly critical, as the output of these agents directly impacts the integrity of the codebase.

In this paper, we bridge this gap by performing an empirical study of guideline documents and review practices at both the project and fine-grained developer levels. We expanded Li et al.'s dataset~\cite{li2025rise} with human-created PRs from the same period, alongside contributor ownership. Our analysis reveals that  AI-assisted PRs have higher percentage of non-owner developers contribution, with over 67.5\% of such PRs originate from contributors without prior code ownership. Yet, the majority (86.9\%) of projects lack guidelines for AI agent usage. Notably, compared to Human-only PRs, Human+AI PRs receive significantly less feedback; contributors with less code ownership are approved faster and with lighter reviews than core members. These findings highlight the need for research into task allocation across human-AI paradigms. Our replication package is available at \url{https://zenodo.org/records/18005310}.

\section{Related Work}
\label{sec:background}
Recent studies show that LLM-based coding tools are already widely used in open-source software (OSS) development \cite{Wang2025}. Tufano et al. \cite{Tufano2024} mined GitHub repositories to identify AI-assisted commits and pull requests, revealing that developers rely on AI tools for tasks including code generation, testing, and documentation. Champa et al. \cite{Champa2024} further showed that developers treat LLMs as collaborative assistants rather than fully autonomous contributors. Similarly, Santos et al. \cite{santos2025} showed that developers use LLMs under human supervision, reporting productivity gains alongside concerns about correctness, limited context awareness, and ethical risks.

Other work has examined how AI-assisted contributions are reviewed and governed in OSS. Chouchen et al.~\cite{Chouchen2024} suggested that AI-involved pull requests tend to be larger and slower to review due to revision overhead. Ogenrwot et al.~\cite{Ogenrwot2024}  found that developers frequently modify or partially reject AI-generated code due to quality and maintainability concerns. Meanwhile, Forsgren et al.~\cite{forsgren2021the} argued that developer productivity and collaboration should be assessed using multiple dimensions, including communication and review interactions, rather than speed alone. This perspective motivates our study, as little is known about differences in review and engagement patterns between AI-assisted and human-created contributions or about the presence of AI usage guidelines.


\section{Study Design and Methodology}
\label{sec:methodology}
We structure our research into three research questions (RQs).

\textbf{RQ1}: To what extent has the reliance on AI-coding agents changed over time, and which types of contributors primarily adopt them? 

\textbf{RQ2}: To what extent is the adoption of AI coding agents planned and discussed, or are there guidelines proposed for their usage?

\textbf{RQ3}: How do community engagement and review practices for AI-assisted contributions differ from those for human-created contributions?

These three RQs progressively investigate the project and contributors' demographics in adopting AI coding agents, relevant guidelines, and the interaction patterns in OSS development.

\subsection{Data Collection}
\label{sec:collection}
To establish a human baseline, we expanded the ``popular'' AIDev dataset by retrieving human-created PRs from the same repositories and timeframes. Specifically, we collected all non-AIDev PRs submitted between the first and last AI-related entry for each repository, yielding 243,306 PRs across 2,253 repositories. We then filtered out bots using Golzadeh et al.'s tool~\cite{Golzadeh2021JSS} and usernames ending in ``bot'' or ``[bot]'', ensuring valid contributors were preserved. The final dataset comprises 174,567 ``Human-only'' PRs across 2,134 repositories, and we make it publicly available.

To mine AI agent usage guidelines, we collected project artefacts reflecting planning or governance decisions. We focused on common files (README, CONTRIBUTING) and, following a manual inspection of 100 repositories, included Code of Conduct and Pull Request templates. We then performed a temporal collection anchored at each repository’s first AI-co-authored PR, capturing an initial version (one month prior) and all updates until the last agentic PR. Table~\ref{tab:artifact_overview} (left) summarises these artefacts, distinguishing pre-adoption guidance from post-adoption updates.

To determine contributors' past ownership in the projects, we adopted the concept of code ownership. Following Lulla et al.~\cite{lulla2025automated}, we used ``Git Blame'' to establish code ownership, as it provides an owner for each line in a file. We calculated the contribution by first checking out the commit immediately preceding the first AI-co-authored PR. Subsequently, we calculated the contribution factors of a contributor across all traceable files as follows: 
 $C_i=\sum_{j\in \mathcal{F}}\frac{l_{ij}}{l_j}$
where $C_i$ stands for the contribution of each contributor, and $\mathcal{F}$ is the set for all files that ``Git Blame'' can trace the contributors.

\subsection{Data Analysis}

To answer RQ1 about the reliance of AI coding agent, we first analysed longitudinal trends by comparing the monthly volume of AI-co-authored and Human-only PRs across active repositories. Subsequently, for each repository, we calculated contributor ownership, ranked developers, and analysed both contributor distribution and PR volume across ranking tiers.

For RQ2 on guidelines, we screen the collected artefacts for AI agent usage guidance using keyword filtering. Specifically, we searched for the five AI agent names, plus ``agent'' and ``agentic''. We analysed full content for initial versions and patches for updates, lowercasing, tokenising, and lemmatising text before matching. Table~\ref{tab:artifact_overview} reports the results. We sampled up to 100 items per artefact for manual annotation, achieving strong agreement (Cohen’s $\kappa$ = 0.85) among two raters \cite{McHugh2012}. We resolved conflicts via discussion to reach consensus. Lastly, we grouped them into themes to investigate their nature.


To answer RQ3 about review practices, we stratified the AIDev dataset into ``AI-only PRs'' and ``Human+AI PRs''. Leveraging the results from our bot-detection process, we identified authors classified as ``bot'', and labelled their contribution as ``AI-only PRs''. The remaining entries in AIDev are labelled as ``Human+AI PRs''. We compared these against the ``Human-only PRs'' from our expanded dataset, excluding any PRs that were not yet closed. These datasets are directly comparable as they cover the same repositories over an identical time period. Furthermore, for ``Human+AI PRs'' and ``Human-only PRs'', we partitioned them based on the creators' ownership. We divided contributors into three cohorts: those ranked in the top 50\% of code ownership, lower 50\% (among contributors with ownership), and those with no prior code ownership, based on the calculation we mentioned in Section~\ref{sec:collection}. These groups are abbreviated as ``Top50\%'', ``Low50\%'', and ``Out'', respectively.

Following this preprocessing, we calculated the average number of comments, code reviews, automated feedback, and time to close for each cohort. We also calculated the total number of PRs that received no feedback other than from AI bots or the PR creators themselves. These metrics serve as a proxy for the review effort spent on different cohorts across three development paradigms.

\begin{table}[]
\centering
\footnotesize
\caption{Dataset overview of collected textual artifacts and keyword-filtered results for RQ2.}
\begin{tabular}{c|rrr|rrr}
\toprule
 & \multicolumn{3}{c}{Data Collection} & \multicolumn{3}{|c}{Keyword Filtering} \\ \cline{2-7}
 & \# Repos & \# Files & \# Patches & \# Repos & \# Files & \# Patches \\
\midrule
Ctb & 1,183 & 1,183 & 350 & 73 & 69 & 15 \\
Rmd & 2,706 & 2,706 & 4,621 & 512 & 478 & 797 \\
CoC & 800 & 800 & 22 & 4 & 3 & 1 \\
PRT & 727 & 727 & 74 & 8 & 8 & 0 \\
\bottomrule 
\multicolumn{7}{l}{\scriptsize Ctb = CONTRIBUTING, Rmd = README, CoC = Code of Conduct, PRT = PR Template.}
\end{tabular}
\label{tab:artifact_overview}
\end{table}

\section{Results and Discussion}
This section contains the result for three RQs and the discussion.
\label{sec:res-disc}

\begin{figure}[h!]
    \centering
    \includegraphics[width=0.75\linewidth]{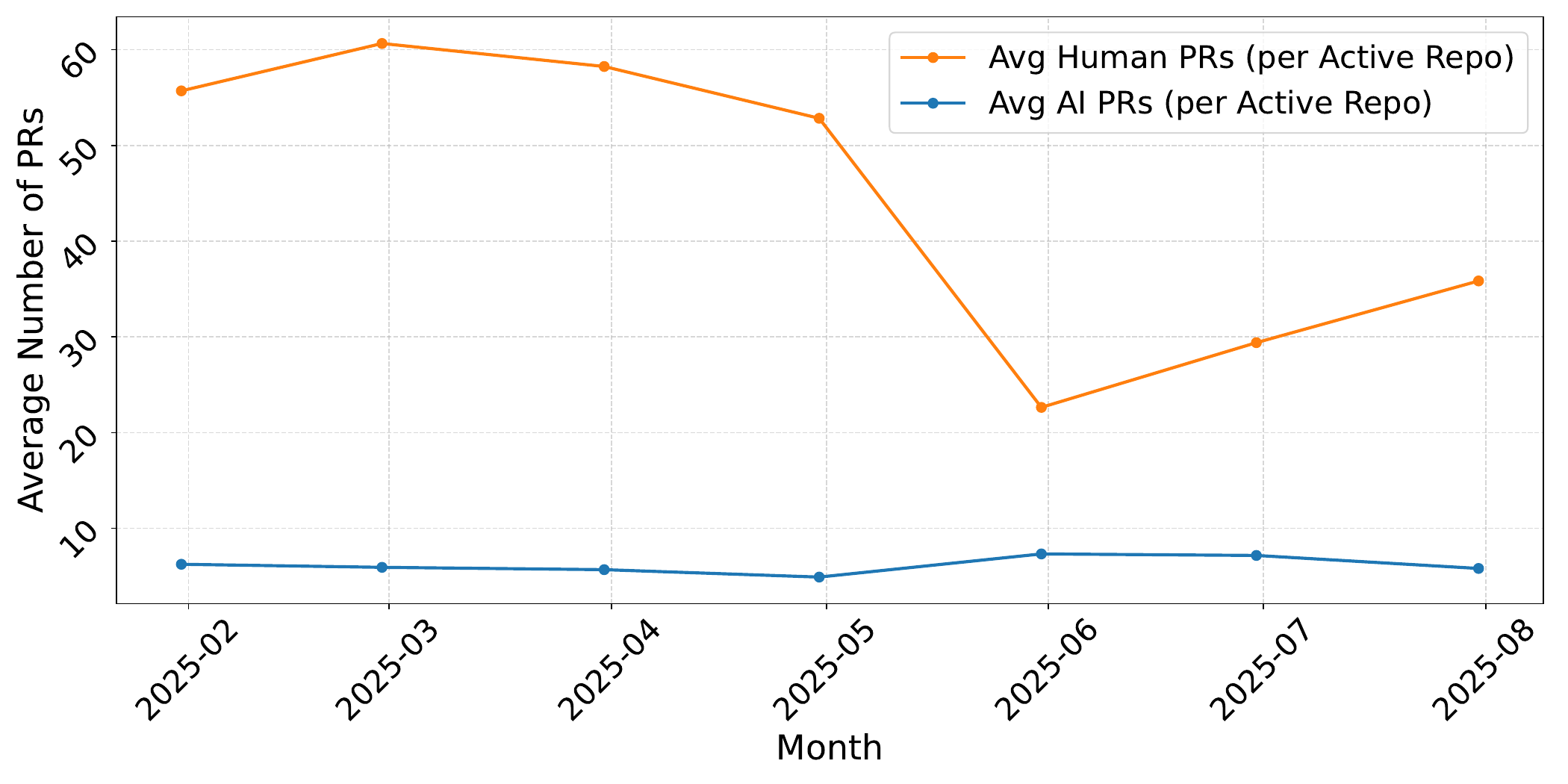}
    \caption{Average Human vs AI PRs across the active months}
    \label{fig:ai_pr_ratio}
\end{figure}

\ul{\textbf{RQ1: AI Coding Agents Adoption.}} First, we present the trends in AI-assisted PR adoption. Figure~\ref{fig:ai_pr_ratio} displays the average monthly volume of AI-assisted versus Human-only PRs. As shown, the frequency of AI-assisted PRs remains low, despite a minor increase since June 2025. However, as more repositories adopt AI agents, the volume of Human-only PRs has declined since June 2025. This divergence has resulted in a higher ratio of AI co-authorship compared to the period preceding this turning point. Next, we analysed contributor cohorts for both types. A comparison of unique counts reveals a significant disparity: while human-created PRs involve 24,000 contributors, AI-assisted PRs originating from the same repositories only account for 1,797 contributors within the same repositories.

\begin{figure*}[h!]
    \centering
    \includegraphics[width=0.65\linewidth]{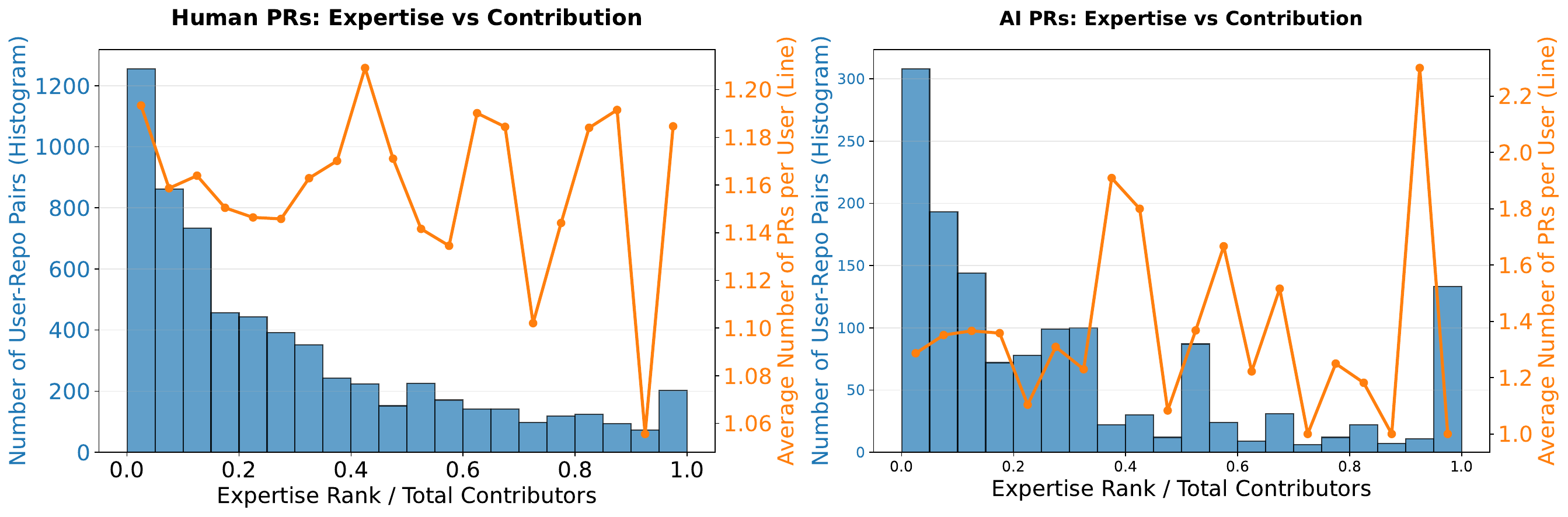}
    \caption{Contributors' Onwership Distribution and Contribution Volume for Human and AI PRs}
    \label{fig:contributor-expertise}
\end{figure*}

Regarding contributor ownership, 32.5\% of AI-assisted PRs (excluding AI-only PRs) involve authors with prior code ownership, compared to only 39.0\% of Human-only PRs. This indicates slightly higher percentage of the AI-assisted PR creators are newcomers to the project, and the vast majority originate from contributors without ownership.  Figure~\ref{fig:contributor-expertise} illustrates the relationship between developers' code ownership and their contribution volume. We defined ownership by rank, where a rank of 1 denotes the contributor with the highest code ownership in the cohort. The data reveals distinct paradigms between human and AI workflows. In the human context, developers with ownership contribute fewer PRs, while the distribution remains diverse, more evenly across the contributors. In contrast, AI-assisted is more more utilised by the top contributors. However, usage is not exclusive to developers with ownership; the majority of users (67.5\%) initiating AI-assisted PRs possess no prior code ownership, suggesting adoption is concentrated at the two ownership extremes.


\begin{customframe}
\textbf{RQ1 Summary:} The recent surge in AI adoption accompanies a decrease in Human-only PRs. Meanwhile, larger proportion of AI-assisted PRs are created by developers without prior code ownership compared Huamn-only PRs (67.5\% vs 61.0\%). Additionally, top-ranking contributors tend to create more AI-assisted PRs.
\end{customframe}

\ul{\textbf{RQ2: AI Coding Agent Guidelines.}}
Following keyword filtering and sampling, we analysed 196 artefacts (116 initial, 80 patches) from 145 repositories. Manual annotation reveals only 21 (10.7\%) contain AI agent guidelines, appearing most commonly in CONTRIBUTING files (13/21, 62.5\%) and PR templates (5/21, 32.8\%). At the repository level, 13.1\% (19/145) possess at least one artefact mentioning AI usage. Among 19 repositories that have the guidelines, 15 (78.9\%) mentioned AI agent usage before their first AI-assisted PR, suggesting planned adoption. Among these, only one repository (6.7\%) updates any related documentation after adoption; however, none introduce new or revised the guidance in later patches. In contrast, four (21.1\%) introduce guidance only after AI-assisted PRs appear, reflecting reactive governance.


To further characterise these 21 guideline items, we grouped them into themes based on the intent and focus of the guidance. We observed that the most common guidance focuses on practical rules for AI contributions. These include disclosure requirements, such as ``prefix your commit message with \texttt{AI:}'' \cite{example1} or ``Please disclose any use of LLMs''~\cite{example2}. Another frequent group provides how-to guidance for AI tools, pointing contributors to rules, prompts, or configurations, for instance ``find additional prompts in .github/prompts or type \#prompt in Copilot'' in \cite{example3}. A smaller but notable group sets boundaries on AI involvement, emphasising human responsibility, for example ``an AI summary isn’t enough by itself'' stated in \cite{example4} or that certain files should be ``written by a human (with minimal help from a language model)'' \cite{example5}. Finally, a few items address community norms, embedding AI use into codes of conduct, for example by giving ``guidelines to follow when using generative AI to answer questions in the community'' \cite{example6}.


\begin{customframe}
\textbf{RQ2 Summary:} Guidance on AI agent usage is rare, indicating that most projects adopt AI agents in an unplanned manner. When present, guidance is typically introduced before the first AI-coauthored PR while a few add guidance only after AI usage begins. Most guidance focuses on disclosure of AI use.
\end{customframe}


\ul{\textbf{RQ3: AI-Human Interaction Patterns.}}
We divided the data into three paradigms: ``AI-only PR'', ``Human+AI PR'', and ``Human-only PRs''. We then divided the latter two based on the contributors' ownership to the project. We further stratified the latter two by contributor ownership. Using prior results, we classified contributors as top-50\% or lower-50\% (among those with ownership), versus those with no ownership. Then we analysed the PR activities for these paradigms.  Table~\ref{tab:result-tab} shows the results.

\begin{table*}[]
    \centering
    \caption{Review Statistics for Different Development Paradigms across Contributors with Different Ownership}
    \resizebox{0.7\linewidth}{!}{%
    \begin{tabular}{l l r r r r r r}
    \toprule
    \multicolumn{2}{c}{\textbf{PR Paradigm}}  & \textbf{\# Entries} & \textbf{Comments} & \textbf{Reviews}  & \textbf{Auto. Feedback} & \textbf{\# No Human} & \textbf{Time to Close} \\
    \midrule
    \multirow{2}{*}{AI-only PR}     & Merged & 4,751 & 3.25 & 3.21 & 1.65 & 83 & 2.42 days \\
    & Unmerged & 3,845 & 2.65 & 0.89  & 1.45 & 31 & 5.53 days \\
    \midrule
    \multirow{2}{*}{Human+AI-Top50\%} & Merged  & 4,953 & 0.68 & 0.62 & 0.18 & 3,004 & 1.03 days\\
    & Unmerged & 1,815 & 0.99 & 0.40 & 0.23 & 842 & 5.39 days \\
    \cmidrule{2-8}
    \multirow{2}{*}{Human+AI-Low50\%} & Merged & 692 & 0.62 & 0.03 & 0.22 & 398 & 0.62 days\\
    & Unmerged & 252 & 0.72 & 0.34 & 0.08 & 129 & 2.26 days\\
    \cmidrule{2-8}
        \multirow{2}{*}{Human+AI-Out} & Merged & 13,618 & 0.18 & 0.15  & 0.05 & 11,817 & 0.18 days\\
    & Unmerged & 2,435 & 0.52 & 0.14 & 0.18 & 1,727 & 2.31 days\\
    \midrule
    \multirow{2}{*}{Human-only-Top50\%} & Merged & 54,261 & 2.58 & 4.98 & 0.53 & 7,971 & 2.32 days \\
    & Unmerged & 6,205 & 3.05 & 4.21  & 0.53 & 1,429 & 15.97 days\\
    \cmidrule{2-8}
    \multirow{2}{*}{Human-only-Low50\%} & Merged & 5,385 & 2.53 & 4.76 & 0.29 & 1,239 & 1.82 days \\
    & Unmerged & 873 & 3.25 & 4.76 & 0.40 & 188 & 17.63 days \\
    \cmidrule{2-8}
    \multirow{2}{*}{Human-only-Out} & Merged & 80,396 & 2.99 & 5.52 & 0.70 & 18,497 & 3.68 days\\
    & Unmerged & 22,130 & 3.77 & 5.11 & 0.69 & 5,823 & 18.86 days\\
    \bottomrule
    \end{tabular}}
    \label{tab:result-tab}
\end{table*}

The table shows Human-only PRs achieved the highest merge rates. Meanwhile, AI-only PRs received only 55.3\%, indicating nearly half were rejected. A closer look at the merge rates for contributors with different levels of ownership reveals opposite patterns for Human+AI and Human-only PRs. For Human-only PRs, the top 50\% contributors received the highest merge rate of 89.7\%, dropping to 78.4\% for non-owners. In contrast, for Human+AI PRs, developers with code ownership received merge rates of 73.2\% and 73.3\% for the top and lower 50\% respectively, while those without code ownership received a highest 84.8\% merge rate.

For AI-only PRs, they took on average roughly the same amount of time to be merged, but much less time to be closed without merging. Only 1.3\% are closed without any comment or review from any contributors in the project, and on average the merged PRs received more comments and reviews by the contributors. For Human+AI PRs, we see significantly less developer effort; the majority (79.0\% merged, 59.9\% closed) received no external reviews or comments. Consequently, the average amount of comments and reviews received are below one per PR for all contributor groups. Still, feedback decreases as contributor ownership lowers. Regarding the average time to close, Human+AI PRs are faster than the other two paradigms, with the merging time goes even below one day for the lower and non-owner developers.

Human-only PRs serve as a baseline of the traditional development patterns in OSS projects. Overall, 19.8\% and 25.5\% of the merged and unmerged PRs are closed without any human comment or review other than the contributor who created the PR. Meanwhile, much higher engagement is witnessed, with merged PRs averaging 4.76 to 5.52 reviews. Contributors who do not have past authorship to the codebase also receive higher amount of reviews and comments. The trend is also in the average time to close a PR. With the average time much higher than Human+AI PRs, and the PRs from contributors who do not have ownership to the codebase also take longer time to be merged.

     

\begin{customframe}
\textbf{RQ3 Summary:} We observe a converse patterns in Human+AI development versus Human-only PRs: contributions from low-ownership developers are approved faster and with less scrutiny than even those from core members.
\end{customframe}

\ul{\textbf{Discussion and Implications.}}  While 67.5\% of Human+AI PRs originate from non-owners, guidelines for AI agents remain scarce, echoing prior guidance gaps~\cite{banyongrakkul2025}. Therefore, similar to other documentation maintenance~\cite{gao2025adapting}, project maintainers should make efforts in regulating the usage of AI coding agent and constantly adapt the instruction for newly-emerged issues.

Our analysis of Human+AI PRs yields counter-intuitive results: most Human+AI PRs are closed without explicit review. Even when reviewed, non-owner contributors tend to receive less feedback compared to core members, which is completely opposite to the pattern observed in Human-only PRs. Confounding factors of this phenomenon could relate to the variability of the development tasks for different development paradigms. Due to space constraints, we did not analyse the reasons behind it. We hypothesise that this may stem from Human+AI PRs primarily targeting low-hanging fruit within the projects. Future research should investigate the types of tasks allocated to contributors with varying ownership across Human+AI, AI-only, and Human-only paradigms. Additionally, studies should examine whether code quality might be compromised due to less rigorous guardrails for Human+AI PRs. Finally, it is worth exploring whether automatic task allocation can be achieved across different development paradigms similar to prior reviewer assignments~\cite{soares2018factors}; specifically for newcomers, researchers could investigate whether the AI-aided assignment of entry-level tasks facilitates better project onboarding~\cite{steinmacher2016overcoming}.

\section{Threats to Validity}
\label{sec:ttv}
Firstly, we assessed developers' profile solely based on code ownership. Incorporating additional signals, such as developer profile metadata or comprehensive commit histories, could provide deeper insights. Second, multiple usernames can link to the same users, and we did not resolve the alias issue~\cite{campanella2024hidden}, which could impact the result. Third, we leveraged the bot-detection algorithm from Golzadeh et al.~\cite{Golzadeh2021JSS}, which relies on comment patterns. Consequently, this approach may fail to identify bots in repositories with limited activity, affecting the precise classification of PRs. Finally, our analysis focused on standard contribution files which might not be comprehensive; future studies could expand this to specialised documentation such as Agent.md~\cite{chatlatanagulchai2025agent}.

\section{Ethical Implications}
\label{ethics}
The code ownership data comprises non-anonymised developer usernames.

\section{Conclusion}
\label{sec:conclusion}
We investigated developer interactions with AI-assisted PRs and project guidelines on them. We mined documentation and expanded the AIDev dataset with contributor ownership and a human baseline. We found that 67.5\% of AI-assisted originate from contributors without code ownership. Yet, 86.9\% of repositories lack guidelines for AI agent usage. Furthermore, Human+AI PRs merge significantly faster with minimal feedback. Notably, non-owner PRs receive the least feedback, with approximately 80\% of merged without explicit reviews, sharply contrasting human PRs where non-owner developers receive the most. We conclude with implications for developers and researchers.

\bibliographystyle{ACM-Reference-Format}
\bibliography{reference}

\end{document}